# Devising Experiments with Interactive Environments

**Pavlos Panagiotidis**
School of Computer Science, University of Nottingham

**Jocelyn Spence**
School of Computer Science, University of Nottingham

**Nils Jäger**
Department of Architecture and Built Environment, Faculty of Engineering, University of Nottingham

## Abstract

This paper reports a practice-based investigation into authoring responsive light and sound in immersive performance without writing code. A modular system couples live gesture, position, and speech inputs to scenographic outputs through a visual logic layer that performers can operate in rehearsal. Across six workshops with eight professional performance-makers, we staged a progression from parallel ensemble and technical training to integrated dramaturgy, culminating in a single-spectator scratch immersive performance with interactive elements. This paper details the system's building blocks and the workshop arc. A reflexive reading of workshop video logs, post-workshop focus groups, and facilitator notes surfaced three ensemble-level strategies that made the technology workable in a hybrid devising/design practice: rotating roles between operator, performer, and mediator; embracing controlled imperfection as a creative resource; and using technology-describing metaphors to support creative practice.

# Introduction

In J.G. Ballard's short sci-fi story *The Thousand Dreams of Stellavista*, houses absorb the emotional residues of their occupants and respond in kind — sometimes violently — offering not only a critique of technological progress but also a provocation to treat architecture as a character shaped by and reactive to human behaviour (Ballard, 1962). This is a useful lens for contemporary immersive performance-making that aims to treat interactive architecture not as static backdrop, but as dramaturgical material in creative processes. In immersive theatrical contexts, the environment itself frames and guides experience in a form of theatrical performance where the world of the event surrounds the audience (Machon, 2013). Warren (2017) further distinguishes between different immersive theatre forms, including *Exploration Theatre*, where audiences may roam freely through space, and *Guided Experiences*, where spatial design channels audience trajectories and choices. In both cases, the built environment can become an active participant in dramaturgy. This aligns with Leatherbarrow's proposition that architecture can be considered less in terms of *what it is* than *what it does* (Leatherbarrow, 2005).

Recent sensing and actuation tools make technologically enhanced, dynamic architectures more accessible than ever. Digital systems using cameras, microphones, and environmental sensors can capture positions, gestures, and voice, while rule-based or AI-driven systems can adapt light, sound, object positions, and other scenographic elements in real time. Yet, utilising this potential requires hybrid processes and methodologies that can bridge design, dramaturgy, and spatial composition in order to align technical responsiveness with embodied creative practice.

This paper reports a practice-based investigation, building on earlier exploratory work that identified the need for rehearsal-friendly, ensemble-accessible approaches to integrating emerging technologies into performance-making. Approaches such as Viewpoints — a compositional and training method that develops ensemble awareness, spatial listening, and improvisational responsiveness (Bogart & Landau, 2005) and Soma Design — an interaction design methodology that grounds technology development in first-person, felt, embodied experience (Höök, 2018) — offered useful scaffolds, foregrounding ensemble awareness, perception-in-action, and embodied experience as bases for performance-making and interaction design. The challenge is not only the choice and configuration of tools, but the framing of space itself as performing agent, with a distinct role and presence.

Through a series of workshops, a group of 8 professional performance-makers integrated a custom-made, modular, no-code interaction authoring workflow, combining computer vision, speech analysis, and visual programming into Viewpoints-based composition making. The workflow (that will be described in detail below) enabled theatre-makers to map embodied inputs (positions, gestures, vocal cues) to scenographic outputs (light and sound) during rehearsal, without writing code or relying on technicians. Our claim is that this combination supports rehearsal-friendly prototyping of responsive environments that treat space as an interactive scenographic layer, while keeping authorship legible to the ensemble.

**System Design Logic**

Accepting that complexity, robustness, and flexibility cannot always be sustained simultaneously, the design of the interaction authoring system adopts a modular design that prioritises rapid, adaptable prototyping to match the tempo and openness of early-stage theatrical devising. Discrete "ingredients", such as gesture recognition, position tracking, speech analysis, sound, and light adaptation, can be swapped or recombined using a visual programming tool. An iterative development cycle followed a hybrid design sprint/devising process: multiple performance prototypes are tested in each workshop, process feedback is gathered immediately, and refinements of the system are rolled out in the following session. Complexity scales with experience, starting from simple interactions (e.g., triggering a single light from a gesture) and building towards compositions, where interactive elements were embedded within short devised scenes and carried specific dramaturgical purposes (e.g., marking an entrance, supporting a revelation, or framing a character interaction). The process culminates in a single-day design/devise sprint producing a scratch interactive performance.

      The system prioritises tracking and adapting features directly relevant to Viewpoints-oriented performance and architectural adaptation. It leverages open-source libraries such as MediaPipe (Lugaresi et al., 2019) for gesture and body tracking, YOLOv8 (Jocher et al., 2023) for human detection, and Vosk (Alpha Cephei, n.d.) for live speech analysis, all integrated through the Node-RED visual programming environment (Node-RED, n.d.). Outputs include light and sound control using consumer-grade hardware, such as Philips Hue smart bulbs (Signify, n.d.). Inputs and outputs communicate via the Open Sound Control (OSC) protocol (Wright & Freed, 1997), enabling users to coordinate conditional interaction flows in real-time to create simple conditional flows: *If [input condition], then [output response]* - [Figures 1-5].

**Figure 1**

*Node-RED interaction flows: gesture→light, position→light, phrase→sound, emotion→sound. Each row in the Node-RED patch represents a simple interaction flow read from left to right: the grey/blue boxes manage OSC communication with the system, the first coloured box defines the trigger condition (e.g., gesture, position, spoken phrase, or emotion), and the second coloured box specifies the resulting action (e.g., activating a light memory or triggering a sound).*

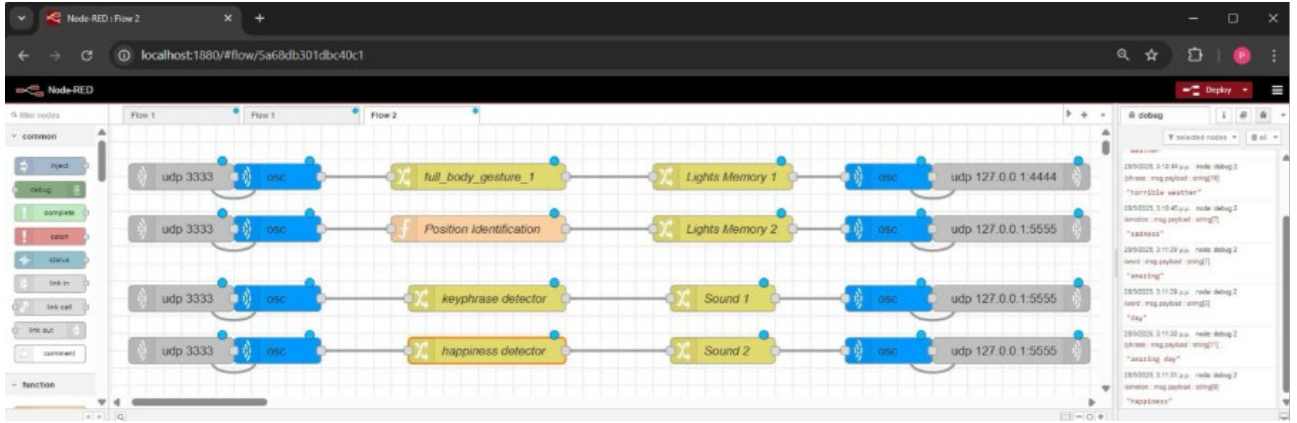

**Figure 2**

*Application of the position tracking script on AI generated image for illustration purposes. YOLOv8 detects human positions from a live feed, calculates pairwise distances, normalises coordinates to a 0–100 scale, and sends both position and distance data via OSC. It logs positions to JSON and generates real-time matplotlib plots for feedback.*

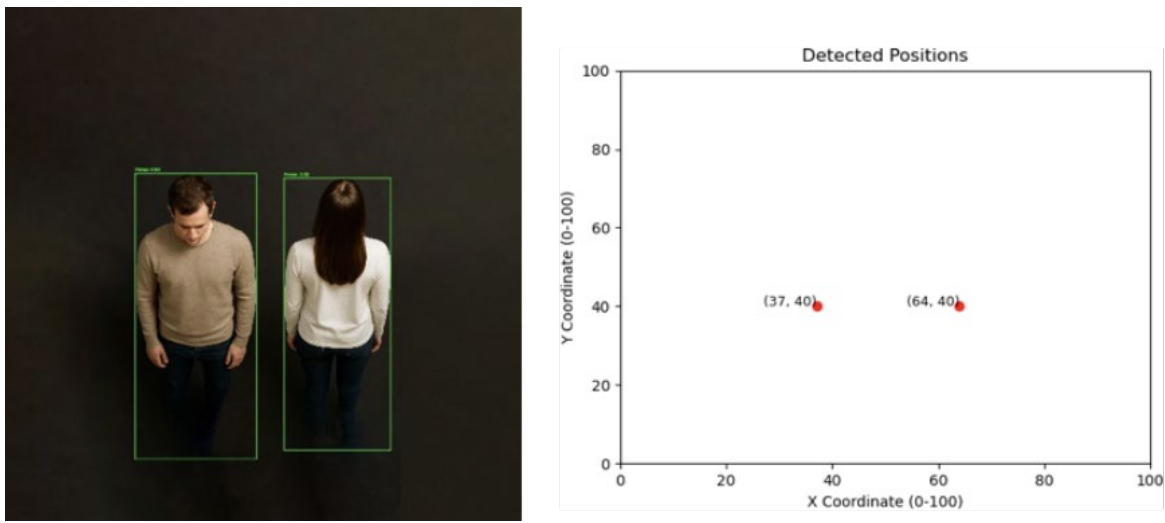

**Figure 3**
*MediaPipe Gesture recognition, gestures are recognised by tracking three points and their angle (left) and Vosk-based speech analysis output (right).*

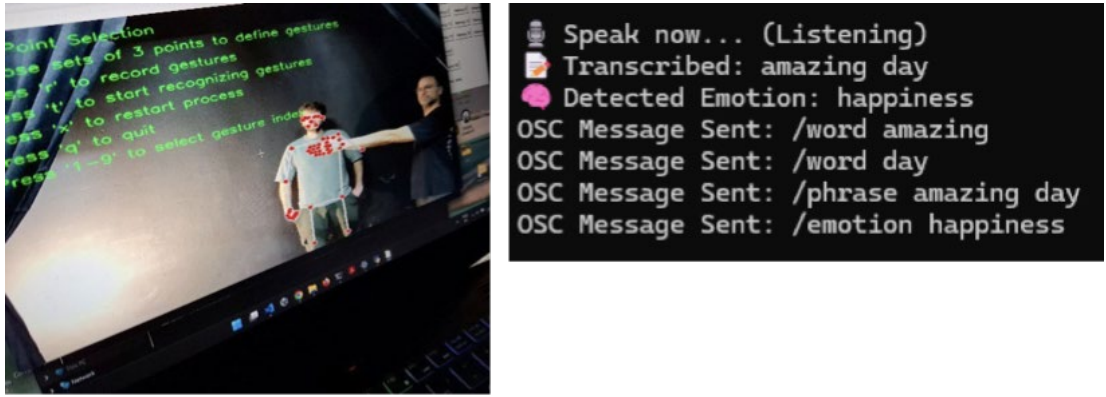

**Figure 4**
*Light control system interface.*
*A custom GUI adjusts brightness, hue, and saturation for Philips Hue lights. Users can create and save/recall up to 20 light configurations ("memories") while they devise the scenes.*

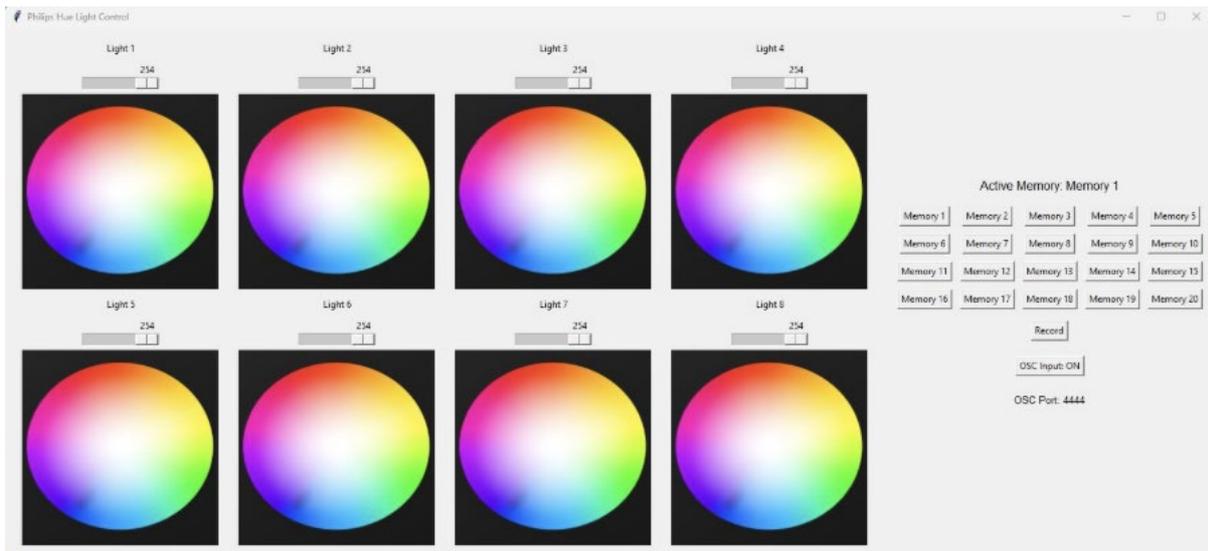

**Figure 5**
*Sound control system interface. A GUI records, loads, and plays up to eight audio clips, assignable to different output devices. Features include a noise gate, device selection, and OSC-triggered playback*

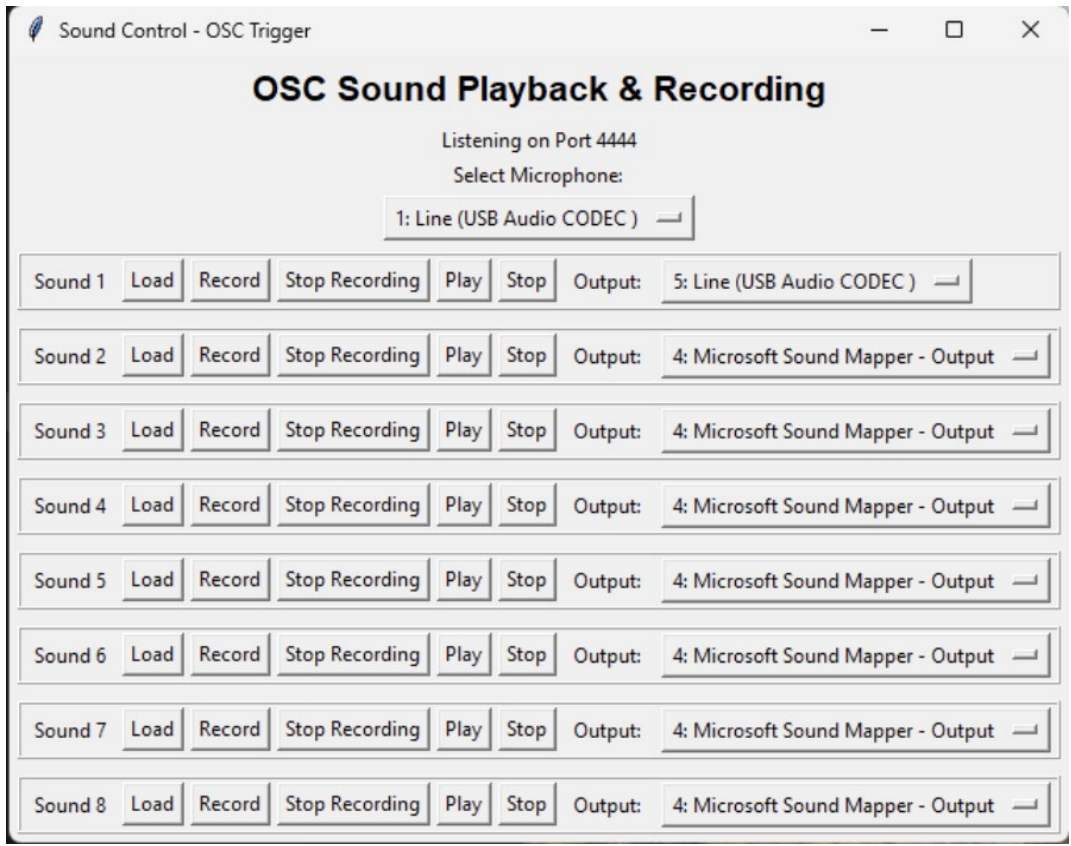

To create a scene, a user can connect these modules through Node-RED. For example: in the Lighting module, define Light Memory 1 so that Lamp A turns red. In the Gesture module, record a gesture such as a raised hand. Then, in Node-RED, link the gesture (input) to Light Memory 1 (output) so that raising the hand turns the lamp red. Similarly, in the Sound module, either record a sound or load a specific audio clip. In the Position module, set the trigger as person detected close to the coordinates (x,y) of the chair. Linking these means that when someone occupies that position, the chosen sound is played. A scene can contain any number of such flows, combining different inputs and outputs as needed (see also Figure 1). In this study architectural adaptations referred mainly to light and sound, but the same no-code system can also drive physical actuators such as motors or linear actuators. This would allow, for example, the opening of doors or movement of panels. Such modules were tested during the system design, though they were not used in the workshops to keep prototyping practical for rapid devising sessions.

**Workshop Implementation: Situated Experiments in Interactivity**

Building on this system logic, the study advanced through six structured workshops. Each was designed as a situated experiment, layering creative and technical complexity step by step while developing familiarity with the responsive authoring tool in real performance contexts. The system designer was also the facilitator that led the workshops, guiding Viewpoints training and technical exercises, while also providing hands-on technical support as needed. The facilitator–designer deliberately refrained from acting as a director during the workshops, only stepping into that role in the final session to link the micro-scenes into a single event. At the outset, the process was deliberately split into two parallel strands: (1) Viewpoints-based ensemble training, which established a shared physical vocabulary, and (2) introductory sessions with the responsive architecture system, where participants experimented with gradually more complex light and sound interactions. This separation aimed to help performers build embodied sensitivity and group cohesion before engaging with the demands of system operation (see Diagram 1).

**Diagram 1**
*Timeline showing the shift from separate training (W1–W2) to discovery (W3), convergence (W4), and integration into dramaturgy and immersive performance (W5–W6).*

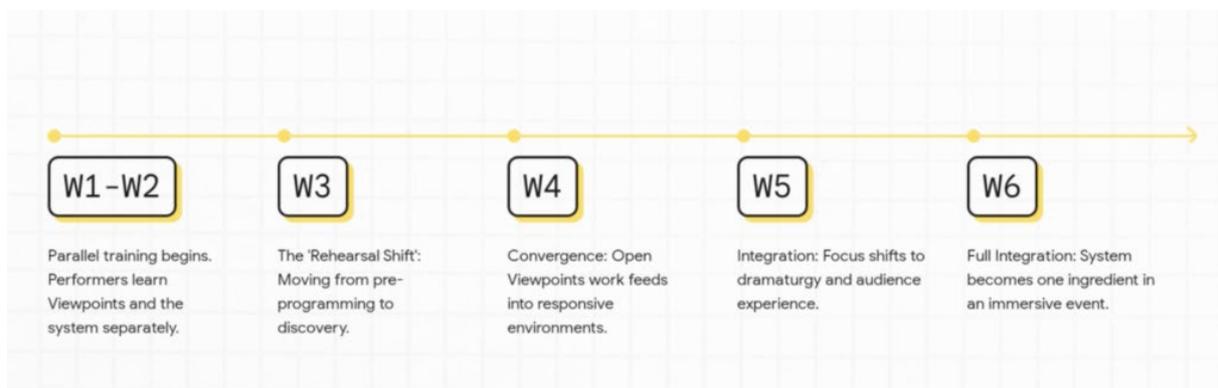

In Workshop 1, participants undertook Viewpoints warm-ups and introductory exercises, followed by basic light control exercises in small groups. This seeded early conversations about automation and distributed control. Workshop 2 bridged into the system's three input modes — gesture, speech, and position — culminating in one-minute scenes per group.  A learning curve emerged, with specific design challenges ('pain points' of the system) such as the need for a more unified interface, occasional tracking inaccuracy and inconsistency, alongside playful 'happy accidents' from unintended interactions.

**Figure 6**

*Left: Improvisation using Viewpoints; Right: Designing Adaptive Spaces.*

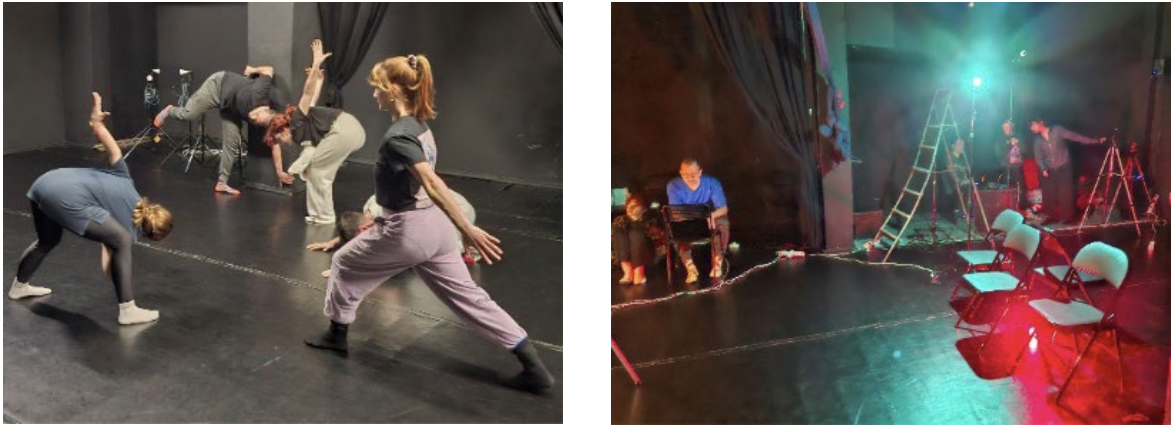

A key shift occurred during Workshop 3, whose focus lay in the groups increasing their collaborative familiarity with the system. This workshop introduced rotation between operator, performer, and mediator roles and a set of supporting documents (cheat sheets, instruction manuals). Additionally, participants moved from attempting to pre-program and execute precise behaviours to discovering architectural responses. The session marked a turning point toward treating the technological activities more as rehearsal rather than technical testing.

In Workshop 4, the two strands converged further: Open Viewpoints work without technology fed directly into improvisations within pre-constructed responsive environments, and then into group-designed setups. Strategic simplification replaced over-complex — and often failed — designs and semi-opaque trigger logic invited participants to experience moments not just of responsiveness but of heightened liveness. As one participant described during the focus group, when lights shifted unexpectedly mid-improvisation, "it was as if you were dealing with something alive…it was very alive in that moment."

**Figure 7**

*Gesture Responsive Lights.*

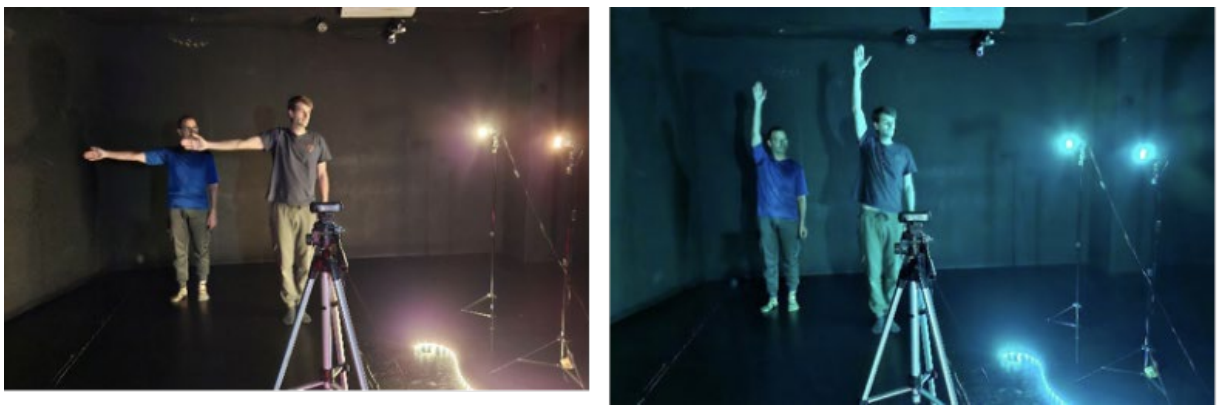

By Workshop 5, the emphasis had rebalanced toward dramaturgy and audience experience. The session began with non-technological Viewpoints to recenter group connection, before devising short participatory micro-performances presented to the rest of the group as audience. Workshop 6 adopted a fast-track scratch performance devising frame — resembling a design sprint — leading to the creation of an immersive event for one audience member. Here, the system was employed selectively and purposefully, with the responsive architecture functioning as one of several performance "ingredients".

**Figure 8**

*W4, Designing Responsive Built Environments.*

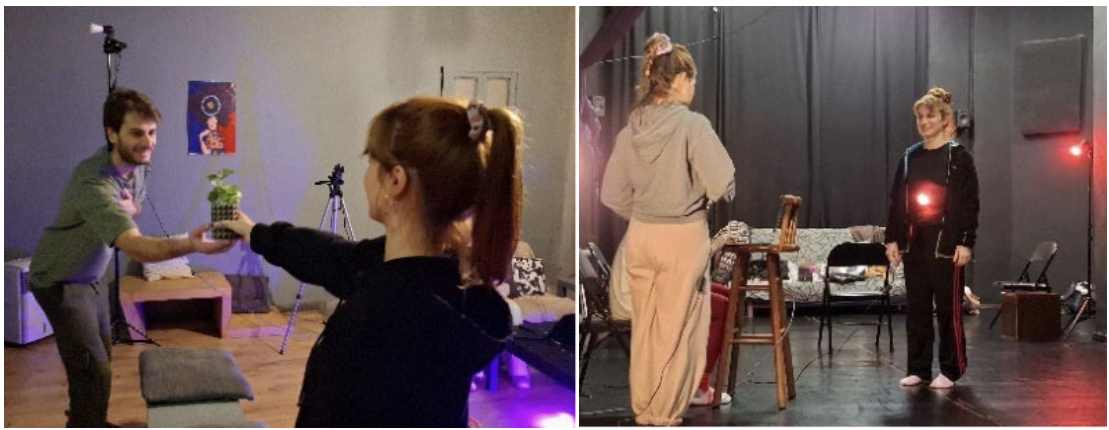

Across the series, the trajectory moved from separation to integration: Workshops 1–3 focused on parallel ensemble and technical training; Workshop 4 moved toward convergence; Workshop 5 embedded responsive design into dramaturgical practice and audience engagement; and Workshop 6 integrated the system fully as one of multiple compositional elements. Each day concluded with a devising session - ranging from one-minute pieces in W1, to participatory micro-performances in W5, and the single-spectator event in W6 — fulfilling the aim that, within a single day, the ensemble could create an immersive scratch composition where interactivity in the built environment was not treated as an external layer managed by a technologist outside the devising group, but was conceived and implemented collectively by the performers themselves within the generative process.

**Figure 9**

*W5, Performers Engaged in Responsive Composition.*

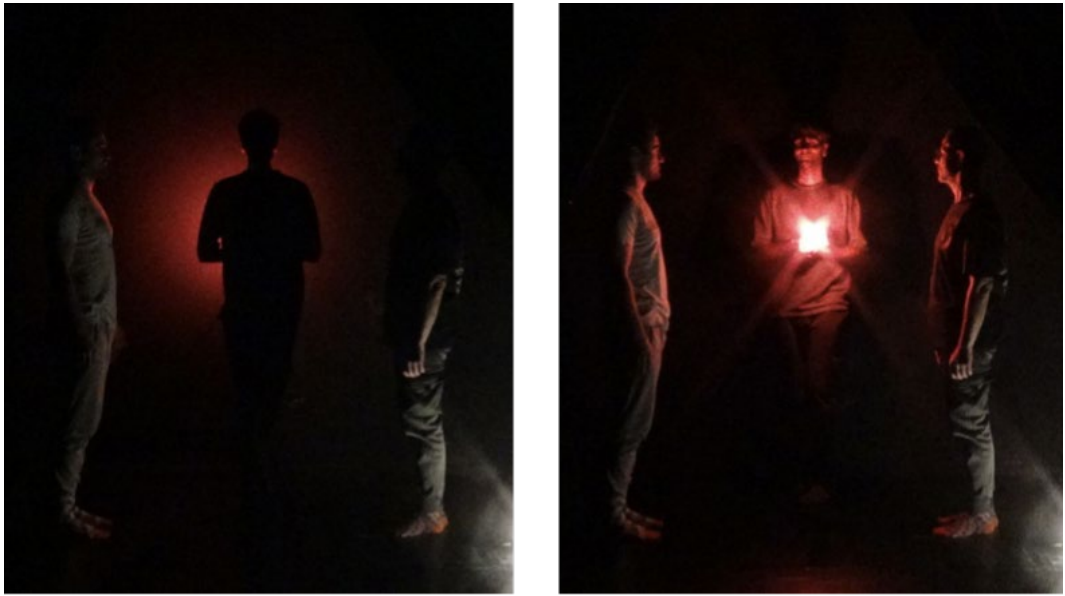

**Figure 10**

*W6, Excerpts from an Immersive Scratch Performance within a Responsive Environment.*

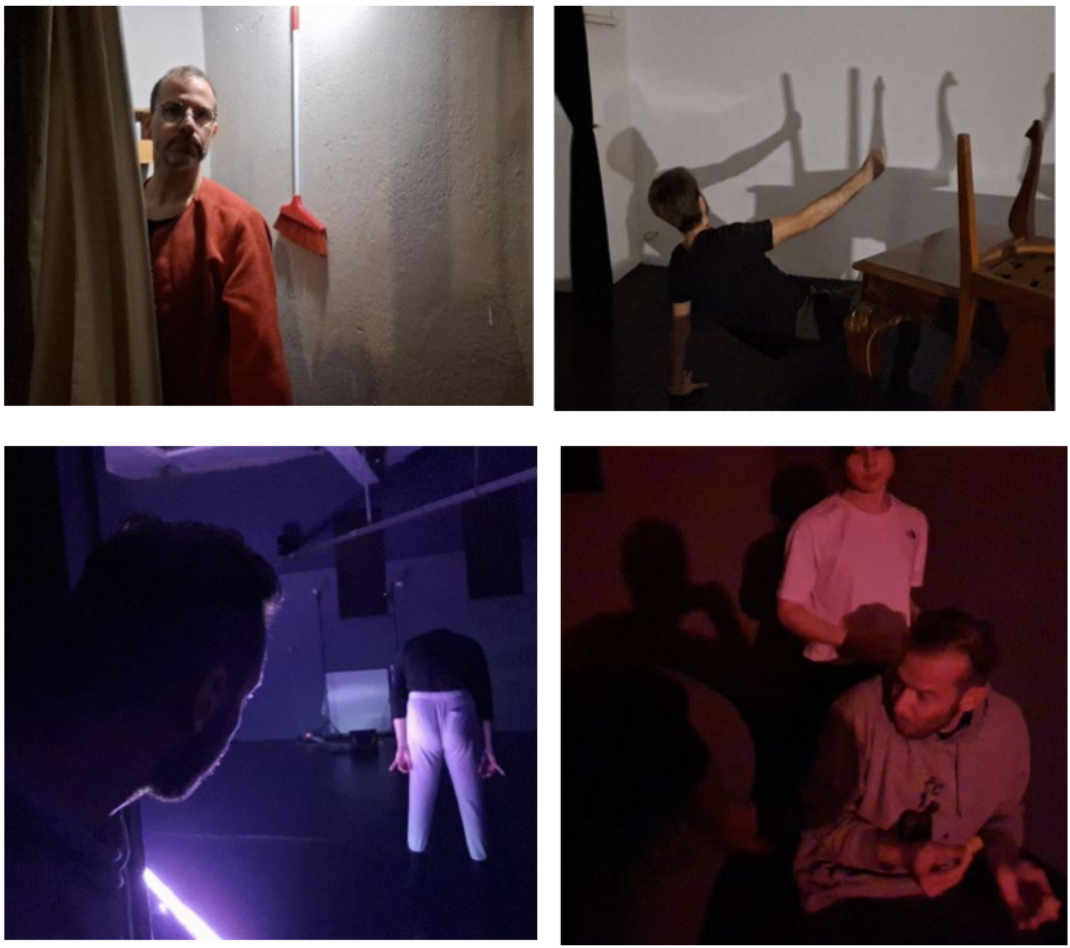

**Emergent patterns and design implications**

This preliminary synthesis draws on reflexive thematic analysis (Braun & Clarke, 2006; 2019) of three primary data sources: (1) video logs coded for learning curves and emergent behaviours; (2) post-workshop focus groups, transcribed and thematically coded; and (3) facilitator field notes capturing design rationale, adjustments, and meta-observations. While full coding tables and longitudinal maps will appear in the extended study, some early tendencies are visible.

Participants engaged with Soma Design concerns by cultivating embodiment and felt experience in interaction with the space. A wake-up stretch, performed as part of a scene, shifted the lights from warm orange to white, evoking sunrise. In this moment the light played the role of the sun, but it was the actor's bodily action that made the sun rise, rather than the other way around. Rotating roles between performer, operator, and mediator distributed technical responsibility and fostered collective ownership. Imperfections in system behaviour were often reframed as generative, with glitches and breakdowns treated as prompts for improvisation. Shared metaphors supported this process, giving the ensemble a vocabulary to negotiate dramaturgy, most notably casting the system as "another performer," a "co-designer," or an "automatic lighting designer."

Together, these tendencies foreground uncertainty in interaction, designing, and engaging with the performing space not as a barrier but as material for collective exploration. They resonate with seamful design in HCI, which treats system imperfections as resources (Chalmers, MacColl & Bell, 2003), and with work on uncomfortable interactions, where disruption can heighten intensity and memorability (Benford et al., 2012). What emerges more distinctly here is how the compositional logic of Viewpoints enables ensembles to absorb such disruptions collectively, shifting emphasis from individual mastery to ensemble adaptability, and positioning responsive systems as potential members of the ensemble.

A preliminary takeaway is that even when interactions are opaque, whether through system glitches, rushed design, or uncertainty about which event triggered a response, they can fuel improvisation and creative exploration. The next step is to examine how such responsive spaces might evolve into full-fledged performance ecologies, and how AI-driven decision engines could further support the role of the environment as an active member of the ensemble.

**Limitations and Conclusions**

This was the first time the responsive architecture system was used in practice, meaning that technical refinement, process design, and creative exploration unfolded in parallel. The workshops were time-compressed, and despite running six sessions, this was not a dedicated user study series but a hybrid of ensemble building, system teaching, iterative system development, and exploratory composition. While the closing sessions demonstrated that participants could create a complete piece mostly independently, some modules — particularly sound and speech — were underused. Certain functions may not have been explained or rehearsed as deeply as they could have been. More time spent solely on system familiarisation might have increased technical fluency, though integrating it with Viewpoints likely helped embed the tool into the ensemble's creative vocabulary, rather than as a separate "tech block." The dual role of system designer as workshop designer/facilitator added significant workload and potential biases in shaping activities and interpreting outcomes.

**Ethics Statement**

Ethical approval was obtained from the University of Nottingham School of Computer Science (Reference: CS-2023-R72), and all participants provided informed consent.

**Acknowledgments**

This work would not have been possible without the commitment and creativity of the eight participating performance-makers — Kornelia Prokopiou, Stylianos Dokouz, Thanos Poumakis, Aspasia Dimou, Pavlina Papadopoulou, Vicky Dalamitra, Orestis Konstantinidis, and Dimitris Koidis — whose openness to experimentation and critical reflection shaped both the process and the outcomes described in this paper. The first author would also like to thank Paul Tennent for his guidance and support throughout this research. This work was supported by the Engineering and Physical Sciences Research Council [Grant number EP/T022493/1], by Lakeside Arts, and by Makers of Imaginary Worlds, as well as by the Horizon Centre for Doctoral Training at the University of Nottingham.